\documentclass[onecolumn,preprintnumbers,amsmath,amssymb,showkeys,12ft]{revtex4}

\usepackage{graphicx}
\usepackage{dcolumn}
\usepackage{bm}

\begin{document}
\draft
\title{Self-organization in nonadditive systems with external noise}
\author{G. Baris Ba\u{g}c\i, Ugur Tirnakli}
\email{ugur.tirnakli@ege.edu.tr}
\address {Department of Physics, Faculty of Science, Ege University, 35100 Izmir, Turkey}

\pagenumbering{arabic}

\begin{abstract}
A nonadditive generalization of Klimontovich's S-theorem [G. B.
Ba\u{g}c\i, Int.J. Mod. Phys. B {\bf 22}, 3381 (2008)] has recently
been obtained by employing Tsallis entropy. This general version
allows one to study physical systems whose stationary distributions
are of the inverse power law in contrast to the original S-theorem,
which only allows exponential stationary distributions. The
nonadditive S-theorem has been applied to the modified Van der Pol
oscillator with inverse power law stationary distribution. By using
nonadditive S-theorem, it is shown that the entropy decreases as the
system is driven out of equilibrium, indicating self-organization in
the system. The allowed values of the nonadditivity index $q$ are
found to be confined to the regime $(0.5,1.0]$.

\end{abstract}

\newpage \setcounter{page}{1}

\maketitle

\section{\protect\bigskip Introduction}

Klimontovich [1] originally proposed S-theorem (Self-organization
theorem) in order to generalize Gibbs' theorem [2] to open systems,
since the latter rests on the assumption that all the compared
distributions have the same mean energy values. However, this
assumption does not hold when one studies open systems with energy
or matter influx. Therefore, Klimontovich considered equating the
mean energies of different states before comparing the associated
entropies. This process of equating mean energies of different
states is called renormalization (not to be confused by
renormalization in quantum field theory) by Klimontovich. However,
S-theorem did not only succeed in generalizing Gibbs' theorem but
gained recognition as a criterion of self-organization in open
systems, since it allows us compare even the stationary
nonequilibrium distributions [3-7].

In open systems, we have some control parameters, which determine
the stationary distributions of the system. As the control parameter
increases, the system recedes away from equilibrium, increasing its
mean energy and entropy. S-theorem orders the associated entropies
in such a way that the state closer to equilibrium possesses a
greater entropy compared to the other states. In other words, we
have a more ordered state as the control parameter increases. This
decrease of entropy on ordering is called self-organization by Haken
[8].

It is worth noting that the use of S-theorem is not limited to
analytical models. It has also been used for many numerical models
such as logistic map [9], heart rate variability [10, 11] and the
analysis of electroencephalograms of epilepsy patients [12] as a
criterion of self-organization.

Despite all its past successes, S-theorem rests on the use of
Boltzmann-Gibbs (BG) entropy. Therefore, it is devised to handle
cases with exponential stationary distributions. On the other hand,
the study of open systems often results in stationary distributions
of the inverse power law form. Hence, it is important to generalize
S-theorem so that its use can be extended to open systems with
stationary distributions of inverse power law form. Such a
generalization has recently been made [13] by the employment of
Tsallis entropy [14-16] instead of BG entropy. However, in Ref.
[13], a rigorous application of the nonadditive S-theorem was not
present [17]. We here present an application of the generalized
S-theorem to the modified Van der Pol oscillator with inverse power
law stationary distributions.

The paper is organized as follows: In Section II, we review the
nonadditive S-theorem for open nonadditive systems. The application
of the nonadditive S-theorem to the modified Van der Pol oscillator
is presented in Section III. Concluding remarks will be presented in
Section IV.

\bigskip

\section{The nonadditive S-theorem}
We begin by supposing that we have two distinct probability
distributions, i.e., $r_{eq}$ and $p$, corresponding to equilibrium
and nonequilibrium states, respectively. In open systems, 
the stationary equilibrium distribution is defined as the distribution 
corresponding to the state where the relevant control parameter 
is set to zero. Similarly, any other stationary state with 
non-zero control parameter is defined as the nonequilibrium state. 
As the value of control parameter increases, the system
recedes away from equilibrium state. The nonadditive S-theorem is
equivalent to showing that the renormalized entropy defined as

\begin{equation}
R_{q}[p\Vert{\widetilde{r}_{eq}}]\equiv\
S_{q}^{neq}(p)-\widetilde{S}_{q}^{eq}(\widetilde{r}_{eq})
\end{equation}

\noindent is negative i.e., $R_{q}<0$, since this implies that
$\widetilde{S}_{q}^{eq}>S_{q}^{neq}$. $S_{q}$ denotes nonadditive
Tsallis entropy

\begin{equation}
S_{q}(p)=\frac{\sum_{i}^{W}p_{i}^{q}-1}{1-q},
\end{equation}

\noindent where $p_{\text{i }}$ is the probability of the system in
the $i$th microstate, $W$ is the total number of the configurations
of the system. The entropic index $q$ is called the nonadditivity
parameter. As $q\rightarrow 1$, the nonadditive Tsallis entropy
becomes

\begin{equation}
S_{q\rightarrow 1}=-\sum_{i=1}^{W}p_{i}\ln p_{i},
\end{equation}

\noindent which is the usual BG entropy.

We denote the equilibrium distribution and the associated entropy in
Eq. (1) by a tilde, since this is not the original equilibrium
entropy but the one obtained after the effective mean energy
equalization i.e., renormalization. This equalization is necessary,
because the mean energies of equilibrium and nonequilibrium states
are different. We define effective mean energy $U_{\text{eff}}$ in terms of
the equilibrium state as

\begin{equation}
U_{\text{eff}}=\ln _{q}\left(\frac{1}{r_{eq}}\right) \; ,
\end{equation}

where the $q$-logarithm function $\ln _{q}(x)$ is defined as

\begin{equation}
\ln _{q}(x)=\frac{x^{1-q}-1}{1-q} .
\end{equation}

\noindent The definition of effective mean energy in Eq. (4) is
central to our generalization and therefore requires some
explanation. The effective mean energy is defined in terms of the
equilibrium distribution obtained from the maximization of Tsallis
entropy subject to ordinary constraints. Therefore, the application
of the effective mean energy definition in Eq. (4) to the
equilibrium distribution $r_{eq}=[1-(q-1)\beta \varepsilon
_{i}]^{1/(q-1)}$ (apart from normalization) results in
$U_{\text{eff}}=\beta\varepsilon_{i}$. This explains why it is called
effective mean energy since it is proportional to the multiplication
of the Lagrange multiplier $\beta$ associated with the internal
energy constraint and the energy of the $i$th microstate. We can
write the equalization of the effective mean energies of two states
as

\begin{equation}
\left\langle U_{\text{eff}}\right\rangle ^{(req)}=\left\langle
U_{\text{eff}}\right\rangle ^{(neq)},
\end{equation}

\noindent where superscripts $(req)$ and $(neq)$ denote the
renormalized equilibrium and ordinary nonequilibrium states,
respectively. From now on, we will drop the subscript $(eq)$ from
the equilibrium probability distribution $r$. Therefore, it should
be understood that the probability distributions $r$ and
$\widetilde{r}$ denote the ordinary and renormalized equilibrium
distributions, respectively. Using normalization and the effective
mean energy definition defined in Eq. (4), Eq. (6) can be written in
a more explicit form as

\begin{equation}
\sum_{i}r_{i}^{q-1}\widetilde{r}_{i}=\sum_{i}r_{i}^{q-1}p_{i},
\end{equation}

\noindent where $\widetilde{r}$ is the renormalized equilibrium
distribution obtained after equating the mean energies.

We then substitute Tsallis entropy given by Eq. (2) into Eq. (1) to
obtain an explicit nonadditive renormalized entropy expression

\begin{equation}
R_{q}[p\Vert\widetilde{r}]=-\left[\frac{1}{(q-1)}\left(\sum_{i}p_{i}^{q}-\sum_{i}
\widetilde{r}_{i}^{q}\right)\right].
\end{equation}

\noindent We can rewrite the above equation as

\bigskip
\begin{equation}
R_{q}[p\Vert\widetilde{r}]=-\left[\frac{1}{(q-1)}\left(\sum_{i}p_{i}^{q}-\sum_{i}
\widetilde{r}_{i}^{q}+(q-1)\sum_{i}\widetilde{r}_{i}^{q}-(q-1)\sum_{i}
\widetilde{r}_{i}^{q}\right)\right].
\end{equation}

\noindent We then substitute $\widetilde{r}_{i}=r_{i}/C$
into Eq. (7) in order to calculate $\sum_{i}\widetilde{r}_{i}^{q}$
explicitly where $C$ is normalization constant. This yields

\begin{equation}
\sum_{i}\widetilde{r}_{i}^{q}=\sum_{i}p_{i}\widetilde{r}_{i}^{q-1}.
\end{equation}

\noindent The substitution of the relation in Eq. (10) into Eq. (9)
finally results

\begin{equation}
R_{q}[p\Vert\widetilde{r}]=-\left(\frac{\sum_{i}p_{i}^{q}}{q-1}+\sum_{i}
\widetilde{r}_{i}^{q}-\frac{1}{q-1}\sum_{i}p_{i}\widetilde{r}_{i}^{q-1}-
\sum_{i}p_{i}\widetilde{r}_{i}^{q-1}\right).
\end{equation}

\noindent In order to consider Eq. (11) as the nonadditive
S-theorem, we must show that the nonadditive renormalized entropy
$R_{q}(p\Vert\widetilde{r})$ is always negative. This can be deduced
from the fact that the expression within the parentheses is the
nonadditive relative entropy $K_{q}[p\Vert \widetilde{r}]$ [18]
i.e.,

\begin{equation}
K_{q}[p\Vert \widetilde{r}]=\frac{\sum_{i}p_{i}^{q}}{q-1}+\sum_{i}\widetilde{r}%
_{i}^{q}-\frac{1}{q-1}\sum_{i}p_{i}\widetilde{r}_{i}^{q-1}-\sum_{i}p_{i}%
\widetilde{r}_{i}^{q-1}.
\end{equation}

\noindent Since the nonadditive relative entropy $K_{q}[p\Vert
\widetilde{r}]$ is positive for all positive $q$ values, we conclude
that

\begin{equation}
R_{q}[p\Vert\widetilde{r}]=S_{q}(p)-\widetilde{S}_{q}(\widetilde{r})=-K_{q}[p\Vert
\widetilde{r}]<0.
\end{equation}

\noindent It is worth noting that the main reason for excluding
negative $q$ values from the above discussion is thermodynamic
stability of Tsallis entropy, since it is stable only for positive
values of $q$ [19].

The nonadditive generalization of S-theorem requires the use of
ordinary probability distributions i.e., first choice of constraints
in nonadditive thermostatistics, instead of escort distributions
i.e., third choice in nonadditive thermostatistics. In fact, even
the nonadditive relative entropy expression we have used is the one
compatible with the ordinary probability distribution [13, 18].

It should be noted that the ordinary S-theorem by Klimontovich is
recovered in the $q\rightarrow1$ limit. This can be seen from the
inspection of Eq. (13), since Tsallis entropy expressions become BG
entropies, whereas all the stationary distributions become
exponential in the $q\rightarrow1$ limit. The nonadditive relative
entropy in Eq. (13) becomes Kullback-Leibler relative entropy in the
aforementioned limit i.e., $K[p\Vert r]\equiv \sum_{i}p_{i}\ln
(p_{i}/r_{i})$ [20], which is positive definite, ensuring the
negativity of the ordinary renormalized entropy.

\section{Noise-driven modified Van der Pol oscillator}
In this section, we will study the noise-driven modified Van der Pol
oscillator, which is classified by the following Ito-Langevin type
two dimensional stochastic equation [21]

\begin{eqnarray}
\dot{x}&=&\frac{\partial H}{\partial y}+\alpha ^{2}f(H;u)\frac{\partial H%
}{\partial x}+\alpha \sigma g(H;u)\xi _{1}(t), \nonumber \\
\dot{y}&=&-\frac{\partial H}{\partial x}+\beta ^{2}f(H;u)\frac{\partial H%
}{\partial y}+\beta \sigma g(H;u)\xi _{2}(t).
\end{eqnarray}

\noindent These type of equations have been first formulated by Enz
[22] in the framework of mixed-canonical-dissipative dynamics and
later used to study nonlinear dynamical systems with noise [23, 24].
The term $H(x,y)$ denotes Hamiltonian. We will particularly consider
the Hamiltonian of the form $H=\frac{1}{2}(x^{2}+y^{2})$, where $x$
and $y$ are fluctuating parameters. $f(H;u)$ and $g(H;u)$ are some
arbitrary functions, which may depend on $H$ and some nonfluctuating
control parameters $u=\{u_{1},u_{2}\}$. Gaussian white noise with
intensity $\sigma =\sqrt{2D}$ is generated through $\xi_{i}$'s,
where $\alpha$ and $\beta$ are real parameters. From now on, we will
assume $\alpha=0$, $\beta=1$, and $g(H;u)=1$. Then, the most general
stationary solution for the modified Van der Pol oscillator in Eq.
(14) is given by

\begin{equation}
f_{0}(H)=C\exp \left[\frac{2}{\sigma ^{2}}\int dHf(H,u_{1},u_{2})\right],
\end{equation}

\noindent where $C$ is normalization constant.

As we remarked above, the function $f(H,u_{1},u_{2})$ is quite
arbitrary and can be changed to another expression as long as the
new expression can be written in terms of the energy $H$ and the
control parameters. We choose the function $f(H,u_{1},u_{2})$ as

\begin{equation}
f(H,u_{1},u_{2})=\frac{-u_{1}-2u_{2}H}{1+\frac{(1-q)}{D}(u_{1}H+u_{2}H^{2})}
\end{equation}
and the control parameters as $u_{1}=a=\gamma-a_{f}$ and
$u_{2}=b/2$, where $\gamma$ is the linear friction coefficient, and
$a_{f}$ is the feedback or control parameter. The term $b$ denotes
the nonlinear friction coefficient. The corresponding stationary
distribution, due to Eq. (15), takes the form

\begin{equation}
f_{0,q}(E)=C\exp_{2-q} \left(-\frac{aE+\frac{1}{2}bE^{2}}{D}\right) \; ,
\end{equation}

\noindent where $q$-exponential is defined by

\begin{equation}
\exp _{q}(x)=[1+(1-q)x]^{\frac{1}{1-q}}.
\end{equation}

\noindent The stationary distribution in Eq. (17) is a
$q$-exponential of the order $(2-q)$, not $q$. In order to
understand this expression, it is important to remember that the
nonadditive S-theorem requires the use of ordinary constraint. The
stationary distribution compatible with the first constraint in
nonadditive thermostatistics is given by $f_{0,q}(E)\propto \exp
_{(2-q)}(-x)$. Following Klimontovich, we assume $b\left\langle
E\right\rangle /\gamma \sim Db/\gamma ^{2}\ll 1$ [4, 13], so that
the corresponding equilibrium distribution associated with zero
value of the control parameter is

\begin{equation}
r_{q}(E)=\frac{\gamma q}{D}\exp_{2-q} \left(-\frac{\gamma E}{D}\right) \; ,
\end{equation}

\noindent where $\gamma q/D$ is the normalization constant
for $q$ values between 0 and 1. Next, we increase the control
parameter to a value different from zero and create nonequilibrium
state in the system. We set the feedback parameter $a_{f}$ equal to
$\gamma$ so that the term $a$ becomes equal to 0. The distribution
function can then be written as

\begin{equation}
p_{q}(E)=\sqrt{\frac{2(1-q)b}{D}}\left[B\left(\frac{1}{2},\frac{1}{1-q}-
\frac{1}{2}\right)\right]^{-1}\left[1+(1-q)\frac{b}{2D}E^{2}\right]^{\frac{1}{q-1}}
\end{equation}

\noindent for $0\leq q\leq1$, where the normalization constant is
inserted [25]. The renormalization of energies i.e., Eq. (7) becomes

\begin{equation}
\int_{0}^{\infty }dEE\widetilde{r}_{q}(E)=\int_{0}^{\infty
}dEEp_{q}(E).
\end{equation}

\noindent The integral on the right hand side can be easily solved
and it yields

\begin{equation}
\int_{0}^{\infty }dEEp_{q}(E)=\sqrt{(1-q)}\left[B\left(\frac{1}{2},\frac{1}{1-q}-
\frac{1}{2}\right)\right]^{-1}\sqrt{\frac{2D}{bq^{2}}}
\end{equation}

\noindent for $0<q\leq1$. On the other hand, the integral on the
left is calculated as

\begin{equation}
\int_{0}^{\infty }dEE\widetilde{r}_{q}(E)=\frac{\widetilde{D}}{(2q-1)\gamma }%
\end{equation}

\noindent for $\frac{1}{2}<q\leq1$. Therefore, Eq. (21),
explicitly written, becomes

\begin{equation}
\widetilde{D}=(2q-1)\gamma \sqrt{(1-q)}\left[B\left(\frac{1}{2},\frac{1}{1-q}-
\frac{1}{2}\right)\right]^{-1}\sqrt{\frac{2D}{bq^{2}}}
\end{equation}

\noindent for $\frac{1}{2}<q\leq1$. Therefore, the nonadditive
renormalized entropy is calculated as

\begin{equation}
R_{q}[p\Vert \widetilde{r}]=(1-q)^{-1}\left[\frac{2(1-q)b}{D}\right]^{(q-1)/2}
\left[B\left(\frac{1}{2},\frac{1}{1-q}-\frac{1}{2}\right)\right]^{-q}
B\left(\frac{1}{2},\frac{q}{1-q}-\frac{1}{2}\right)-
(1-q)^{-1}q^{q}\gamma ^{q-1}\frac{\widetilde{D}^{1-q}}{(2q-1)}
\end{equation}

\noindent for $\frac{1}{2}<q\leq1$, where the renormalized noise
intensity $\widetilde{D}$ is given by Eq. (24). The nonadditive
renormalized entropy expression above is always negative for
positive values of the nonadditivity parameter $q$ i.e.,
$R_{q}[p\Vert\widetilde{r}]<0$, since it is the negative of the
nonadditive relative entropy $K_{q}[p\Vert \widetilde{r}]$ together
with the renormalization of the corresponding mean energies. The
nonadditive renormalized entropy in Eq. (25) is plotted in Figs. 1
and 2 for some particular values of the intensity of the random
source $D$ and nonlinear friction coefficient $b$. However, it is
important to understand that our results are independent of the
random source intensity and nonlinear friction coefficient. The
nonadditive renormalized entropy attains the value $-0.05$ in the
$q\rightarrow1$ limit. This is the exact value one would obtain if
one would use Klimontovich's additive renormalized entropy as a
criterion of self-organization.

\begin{figure}
 \begin{center}
 \includegraphics[scale = 0.8]{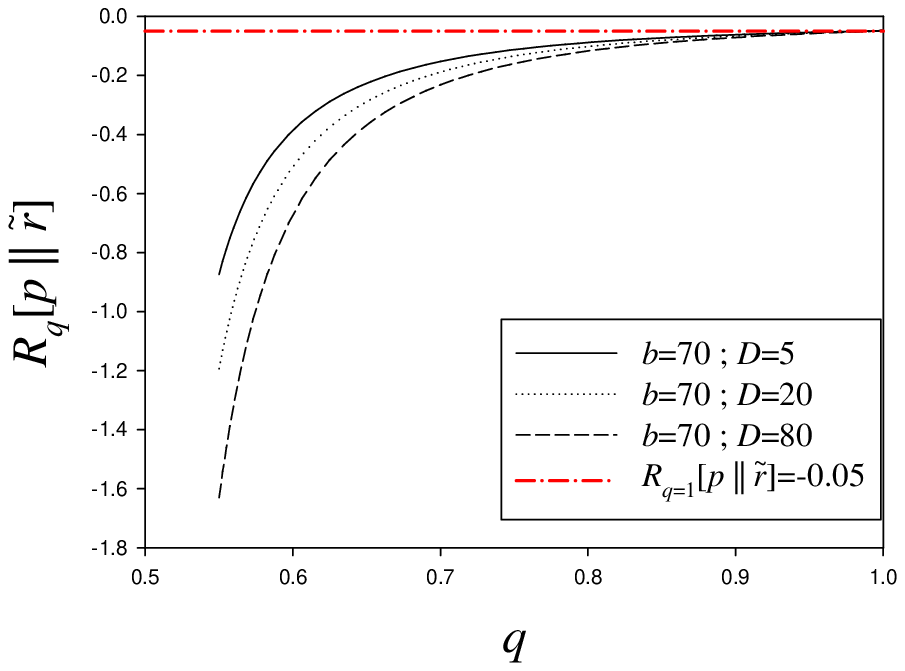}
 \caption{The plot of nonadditive renormalized entropy $R_{q}[p\Vert\widetilde{r}]$ 
versus the nonadditivity parameter $q$ for three representative values of random 
source intensity $D$, where nonlinear friction coefficient is $b$=70.}
 \end{center}
\end{figure}

\begin{figure}
 \begin{center}
 \includegraphics[scale = 0.8]{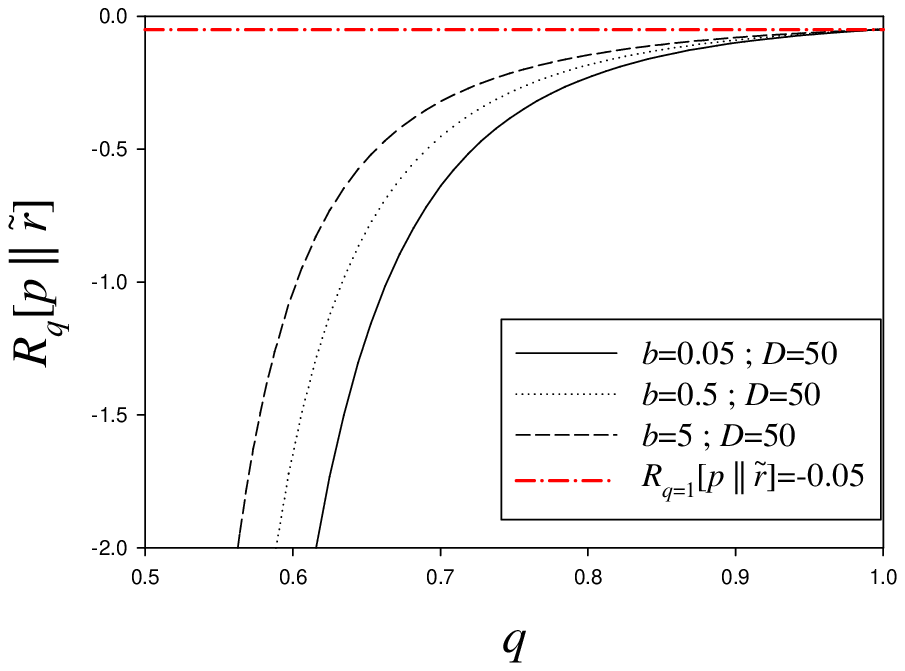}
\caption{The plot of nonadditive renormalized entropy $R_{q}[p\Vert\widetilde{r}]$ 
versus the nonadditivity parameter $q$ for three representative values of nonlinear 
friction coefficient $b$, where random source intensity $D$=50.}
 \end{center}
\end{figure}

\section{Conclusions}
Klimontovich's original S-theorem is based on turning the ordinary
reference distribution into the corresponding escort distribution
where the exponent of the escort distribution is fixed through mean
energy renormalization [3]. However, due to the use of BG entropy,
original S-theorem is limited to the additive open systems, and
exponential stationary distributions. In contrast, the nonadditive
S-theorem is obtained only by using ordinary probability
distributions. When we adopt nonadditive Tsallis entropy, we are
required to use ordinary probability instead of escort distribution,
since the nonadditivity index $q$ plays the role of the escort
distribution's exponent [13].

Klimontovich applied his original S-theorem to Van der Pol
oscillator with exponential stationary distributions [26]. In this
work, the nonadditive S-theorem has been applied to a modified Van
der Pol oscillator with stationary distributions of inverse power
law. It is interesting to note that we have modified Van der Pol
model of Klimontovich only by changing the dissipative term
$f(H,u_{1},u_{2})$. We have shown that the
self-organization takes place as the control parameter increases
from zero value, where the signature of the self-organization in 
this framework is the negativity of the nonadditive renormalized entropy. 
The nonadditive renormalized entropy attains the value $-0.05$ in
the $q\rightarrow1$ limit. This is exactly the same value one would
obtain by using Klimontovich's original S-theorem [4, 13].

The most important difference in changing the underlying dynamics so
that the stationary distribution is of the q-exponential form is the
confinement of the $q$ values to a range between 0.5 and 1. If one
uses nonadditive renormalized entropy together with the exponential
stationary distributions obtained from Klimontovich's Van der Pol
model, the nonadditivity index $q$ can take any value without
limitations [13]. However, the change in the dynamics so as to make
the stationary distributions of the Van der Pol model of
q-exponential form, results in a spectrum of some privileged $q$
values. This seems to be a general feature one encounters whenever
one studies physical systems depending on a control parameter [27,
28]. It is also worth remarking how the probability distribution of the
first constraint in nonadditive thermostatistics i.e.,
$q$-exponential with order $(2-q)$, emerges in the numerical studies
concerning systems depending upon control parameters [27, 28].

\section{ACKNOWLEDGEMENTS}
We thank B. Bakar for his helpful comments. 
This work has been supported by TUBITAK (Turkish Agency) under the 
Research Project number 108T013.



\begin{references}

\bibitem{R1}  Yu. L. Klimontovich, Z. Phys. B {\bf 66}, 125 (1987).

\bibitem{R2}  J. W. Gibbs, Elementary Principles in Statistical Mechanics, Ox Bow Press, New York, 1992.

\bibitem{R3}  Yu. L. Klimontovich, Chaos, Solitons and Fractals {\bf 5}, 1985 (1994).

\bibitem{R4}  Yu. L. Klimontovich, Physica A {\bf 142}, 390 (1987).

\bibitem{R5}  Yu. L. Klimontovich, Turbulent Motion and the Structure of Chaos: A New Approach to the Statistical Theory of Open System, Kluwer Academic Publishers, Dordrecht, 1991.

\bibitem{R6}  Yu. L. Klimontovich, Physica Scripta {\bf 58}, 549 (1998).

\bibitem{R7}  Yu. L. Klimontovich and M. Bonitz, Z. Phys. B {\bf 70}, 241 (1988).

\bibitem{R8}  H. Haken, Information and Self-organization: A Macroscopic Approach to Complex Systems, Springer Verlag, Berlin, 2000.

\bibitem{R9}  P. Saparin, A. Witt, J. Kurths, V. Anischenko, Chaos, Solitons and Fractals{\bf 4}, 1907 (1994).

\bibitem{R10}  J. Kurths \textit{et al.}, Chaos {\bf 5}, 88 (1995).

\bibitem{R11}  A. Voss \textit{et al.}, Cardiovasc. Res. {\bf 31}, 419 (1996).

\bibitem{R12}  K. Kopitzki, P. C. Warnke, J. Timmer, Phys. Rev. E {\bf 58}, 4859 (1998).

\bibitem{R13}  G. B. Ba\u{g}c\i, Int.J. Mod. Phys. B {\bf 22}, 3381 (2008).

\bibitem{R14}  C. Tsallis, J.Stat. Phys. {\bf 52}, 479 (1988).

\bibitem{R15}  E.M.F. Curado and C. Tsallis, J. Phys. A {\bf 24} (1991) L69;
Corrigenda: J. Phys. A {\bf 24} (1991) 3187; {\bf 25}, 1019 (1992).

\bibitem{R16}  C. Tsallis, R. S. Mendes, and A. R. Plastino, Physica A {\bf 261}, 534 (1998).

\bibitem{R17}  In Ref. [13], the nonadditive S-theorem has been applied to the Van
der Pol oscillator model of Klimontovich. However, the stationary
distributions in that model were of exponential form. Therefore,
although important for illustrative reasons, it cannot be considered
as a rigorous application of the nonadditive S-theorem, since the
nonadditive S-theorem relies on the use of Tsallis entropy, which
requires the use of inverse power law stationary distributions.

\bibitem{R18}  S. Abe, G. B. Ba\u{g}c\i, Phys. Rev. E {\bf 71}, 016139 (2005).

\bibitem{R19}  S. Abe, Phys. Rev. E {\bf 66}, 046134 (2002).

\bibitem{R20}  R. Gray, Entropy and Information Entropy, Springer-Verlag, New York, 1990.

\bibitem{R21}  M. Akimoto and A. Suzuki, AIP Conference Proceedings {\bf 708}, 794 (2004).

\bibitem{R22}  C. P. Enz, Physica A {\bf 89}, 1 (1977).

\bibitem{R23}  M-O. Hongler and D. M. Ryter, Z. Phys. B {\bf 31}, 333 (1978).

\bibitem{R24}  W. Ebeling and H. Engel-Herbert, Physica A {\bf 104}, 378 (1980).

\bibitem{R25}  I. S Gradshteyn and I. M. Ryzhik, Table of Integrals, Series, and Products, Academic Press, California, 2000.

\bibitem{R26}  Although Klimontovich calls his model as Van der Pol oscillator,
Klimontovich's Van der Pol model is already a modified one, since it
is different than the original Van der Pol model by an extra $v^{3}$ term.
In this sense, our model can aptly be called a second modified Van
der Pol model, compared to Klimontovich's one.


\bibitem{R27}  A. Robledo, Physica A {\bf 342}, 104 (2004).

\bibitem{R28}  A. Robledo, Physica A {\bf 370}, 449 (2006).


\end{references}
\end{document}